\documentclass[12pt]{article}
\usepackage{graphicx,amsmath,amssymb,amsthm,euscript}
\newcommand{\ssy}[5]{#1,  #2  {\bf #3}, #5 (#4)\rlap{.}}
\newcommand{\rmd}{\mathrm{d}}
\newcommand{\shir}{ w}
\DeclareMathOperator{\Bd}{Bd}
\title{Even the Minkowski space is holed}
\author{S. Krasnikov}
\date{}
\begin{document}
\maketitle
\begin{abstract}
To cure the lack of predictive power of general relativity, Geroch
proposed to complete the theory with an additional postulate that only
``hole-free" spacetimes are permitted. This postulate (or, rather, that
obtained from it by some disambiguating) seems physically well-founded
and at the same time appropriately restrictive. I show, however, that it
is too strong --- it prohibits even the Minkowski space.
\end{abstract}
\section{Introduction}
\begin{figure}[htb]\begin{center}
\includegraphics[width=\textwidth]{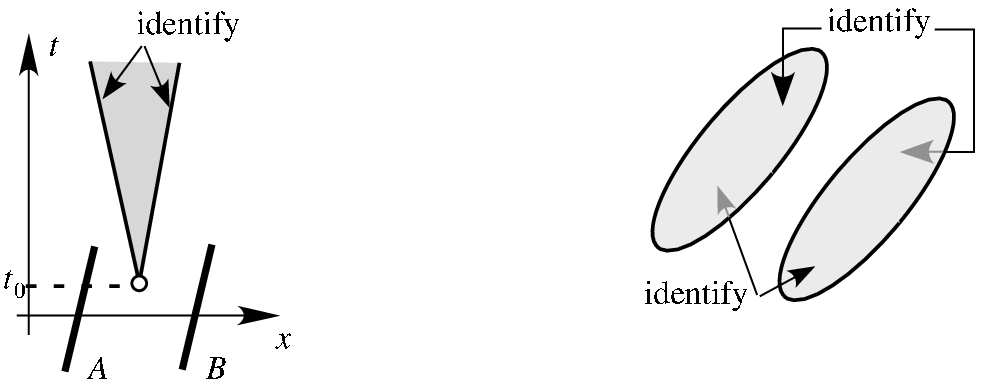}\\
(a)\hspace{0.7\textwidth} (b)
\end{center}\caption{(a) The spacetime is obtained by cutting out the
shadowed sector from the Minkowski space and gluing together the rays
bounding the sector (the rays  do not comprise the vertex of the angle).
(b) Remove a disc $D$ from a domain $N$\label{fig:primery}, take a copy
of this incised spacetime, and identify the upper bank of either cut with
the lower bank of the other. If $N$ and $N'$ are different spacetimes,
the result is a double covering of $N-S$, where $S=\Bd D$ is a sphere of
co-dimension 2. If they are just different regions of the same spacetime,
then, depending on the position of the disks, the resulting spacetime is
either a time machine
\cite{Deu} or a wormhole \cite{viss}.}
\end{figure}

Consider a space obtained by removing from the Minkowski plane the angle
$$|t/x|>2,$$ [see Fig.~\ref{fig:primery}(a)], and by gluing then the rays
$t/x=\pm 2,\ t>0$ together (the points are identified if they have the
same $t$-coordinate). Evidently, $M$, though being a legitimate
spacetime\footnote{It is a smooth connected pseudo-Riemannian manifold.},
is singular and it is the singularities of this type that are our subject
(a simple example of a non-flat spacetime $M$ with a similar singularity
is this: from an arbitrary $n$-dimensional spacetime remove an
$(n-2)$-dimensional sphere and let $M$ be the double covering of the
resulting space, see Fig.~\ref{fig:primery}b). Their peculiarity is that
they, in fact, deprive general relativity of its predictive power.
Indeed, in contrast to the ``usual", curvature singularities, these
``topological" ones are absolutely ``sudden":
\emph{nothing} would tell an observer approaching such a singularity that
his world line will terminate in a moment. In the presence of such
singularities, everything (the geometry of the universe, its topology,
causal structure, etc.) may change whimsically  and (apparently)
causelessly. For example, the spacetime depicted in
Fig.~\ref{fig:primery}(a)
is just the Minkowski space up to some moment $t_0$. But after that
moment some observers ($A$ and $B$ in the figure) will discover that
without experiencing any acceleration they started to move towards each
other. Figure~\ref{fig:primery}(b) shows how in the otherwise Minkowskian
space a time machine or a wormhole may appear with no visible cause (more
of bizarre examples can be found in \cite{Str}). ``Thus general
relativity, which seemed at first as though it would admit a natural and
powerful statement at prediction, apparently does not"
\cite{GerHole}.

The real problem with the just described singularities is that none of
them have ever been observed (see \cite{GRStr}, though). So, it seems
that we are overlooking some fundamental law of nature. Arguablyî this
law may be formulated as an additional (non-local, of course) requirement
on the structure of spacetime which would explicitly prohibit some
singularities. In looking for such a requirement it is desirable, first,
to comprehend what
\emph{exactly} is to be prohibited. In particular, each of the singular
spacetimes mentioned above can be viewed as a quasiregular singularity
(i.~e., a singularity with bounded curvature, see
\cite{quasireg} for a rigorous definition), or as an ``absolutely mild
singularity"  \cite{Str} (it has a finite covering by open sets, each of
which can be extended to a singularity-free spacetime). It is also a
``locally extendible" spacetime (i.~e., it contains an open set $U\subset
M$ isometric  to a subset $U'$ of some other spacetime $M'$ such that the
closure of $U'$ in $M'$ is compact while the closure of $U$ in $M$ is not
\cite{HawEl}). And, finally, it is a ``hole'' in the following sense (up
to two subtle points, the definition below is that of \cite{GerHole}; the
difference is brought about by disambiguation of the latter, see the
Remark below).
\paragraph{Definition.} Denote by $D^+(S)$  the collection
of all points $p$ of $M$ such that every future-directed timelike curve
in $M$, having future endpoint $p$ and no past endpoint, meets $S\subset
M$. A space-time $(M, g)$ is called \emph{hole-free} if it has the
following property: given any achronal hypersurface\footnote{In Geroch's
paper, in contrast to this one, the four-dimensional  case is considered,
but the generalization of what follows to four dimensions is trivial.}
$\EuScript S$ in $M$ and any metric preserving embedding $\pi$ of an open
neighbourhood $U$ of $D^+(\EuScript S)$ into some other spacetime
$(M',g')$, then $\pi(D^+(\EuScript S))= D^+(\pi(\EuScript S))$.

\medskip

\noindent So, which of these | quite different | classes of spacetimes
should be excluded?

Two first possibilities seem too \emph{ad hoc}. The third one once seemed
more promising, but Beem and Ehrlich demonstrated \cite{BeE} that even
the Minkowski space is locally extendible | which, in my belief, rules
out this variant: a postulate is definitely \emph{too} strong if it
excludes the Minkowski space. So, we are left\footnote{As long, that is,
as we are restricted to the classes listed above. There are also
holes$^*$
\cite{Manchak}, Clarke holes \cite{Clarke}, and  more.} with
Geroch's proposal to ``modify general relativity as follows: the new
theory is to be general relativity, but with the additional condition
that only hole-free spacetimes are permitted''. The proposal seems to be
physically well motivated, the idea behind it being to remedy the
following ``defect" of general relativity: ``\dots although $S$
determines what happens in $D^+(S)$, what this $D^+(S)$ will be, and in
particular how ``large" it is, requires knowledge not only of $S$, but
also of the space-time $M$, $g$ in which $S$ is embedded"
\cite{GerHole}.

Recently,  Manchak has constructed a spacetime which is inextendible and
globally hyperbolic but fails to be hole-free
\cite{Manchak}. This spacetime possesses a nasty singularity, so one
might wonder if its exclusion is that great loss for the theory. Still,
some suspicion appeared that the requirement for a physically reasonable
spacetime to be hole-free might be too strong. In the present paper I
show that this is the case.

\paragraph{Remarks on terminology.} 1) In giving the definition to $D^+(S)$,
Geroch \cite{GerHole} refers to \cite{HawEl}, where that object is
defined for \emph{arbitrary} $S$. On the other hand,  \emph{explicitly}
the   definition in
\cite{GerHole} is formulated only for sets $S$ which are
three-dimensional achronal surfaces and this is in accord with
\cite{Ger:domain}, to which a reference is given, too. This discrepancy
is immaterial  within the range of problems studied in
\cite{Ger:domain}\footnote{``[\dots] the definition of the domain of
dependence could be applied equally well when $S$ is not achronal, but
[\dots] the consequence is to introduce additional complications without
adding anything really new" \cite{Ger:domain}.}, but becomes important in
our case: in defining hole-freeness one could | without contradicting
\cite{GerHole}, formally at least | restrict oneself to $\EuScript S$ and
$\pi$
such that $\pi(\EuScript S)$ is achronal too. One would then arrive at a
\emph{different} property | let us call it
\emph{hole-freeness$'$} | and, correspondingly, at the postulate that only
hole-free$'$ spacetimes are permitted\footnote{It is probably this
approach that eventually leads to the concept of
hole-freeness$^*$~\cite{Manchak}.}. This postulate is less restrictive
and fails to fulfill its function:  the existence of a point is now
determined not only by $\EuScript S$, but also by the entire
$I^-(\EuScript S)$, so by requiring that spacetimes would be hole-free$'$
one does not eliminate the above-mentioned ``defect". Which suggests that
hole-freeness in Geroch's proposal must be understood in the sense of the
above-formulated definition. 2) We have chosen the domain of $\pi$ to be
an open neighbourhood of $D^+(\EuScript S)$ [not just $D^+(\EuScript S)$,
as in \cite{GerHole}] to avoid some ambiguity arising when  the term
``embedding" is applied to a set which is not a manifold. This must not,
however, lead to any weakening of the result, because a metric preserving
embedding of $U$ is at the same time a metric preserving embedding of
$D^+(\EuScript S)$ whichever way the latter is defined.

\paragraph{Proposition.} The Minkowski plane is not hole-free.
\section{Proof}
\begin{figure}[htb]\begin{center}
\includegraphics[width=\textwidth]{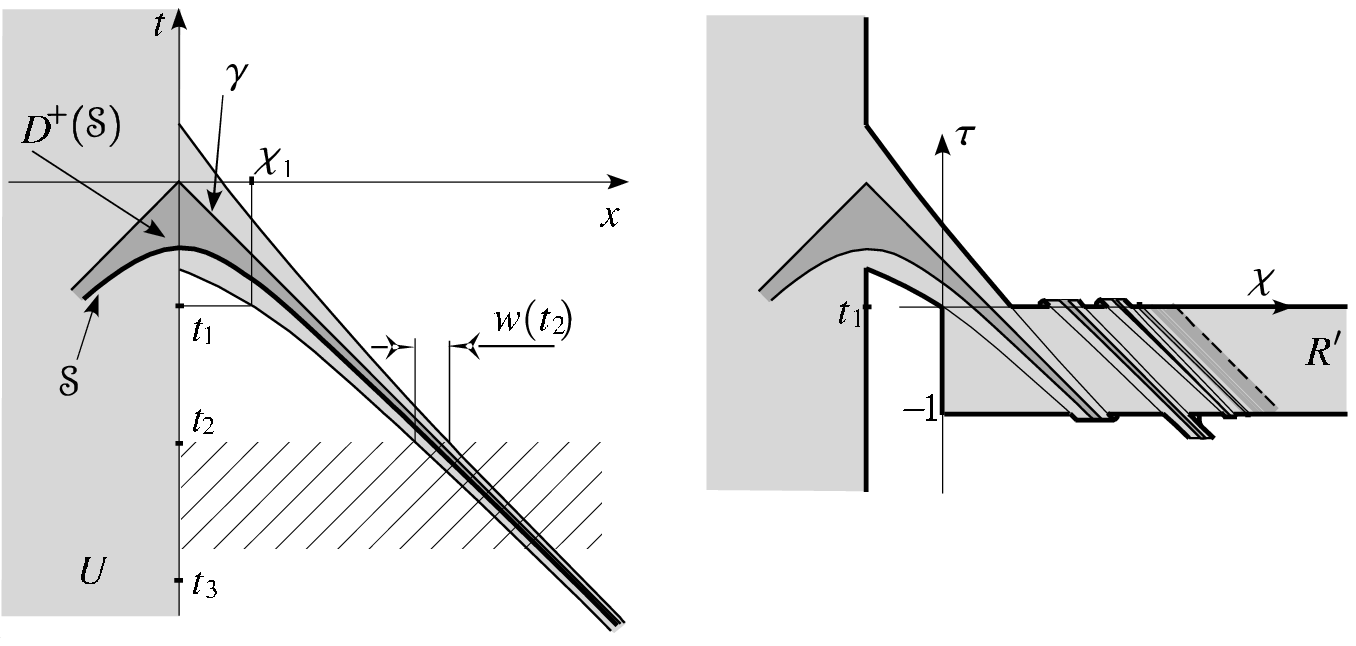}\\
(a)\hspace{0.7\textwidth} (b)
\end{center}\caption{(a) The light-gray area $U$ is a neighbourhood of
the dark-gray area \label{fig:Ah}$D^+(\EuScript S)$. The intersection of
$U$ with the hatched strip is $O_2$. (b) The light-gray area is $M'$ | an
extension of $U$ alternative to the Minkowski plane.}\end{figure}

Let $\EuScript S$ be the   hyperbolae $t=-\sqrt{x^2+1}$. Then obviously
$D^+(\EuScript S)$ is the closed set bounded by this hyperbolae from
below and by the angle $t=-|x|$ from above, see figure~\ref{fig:Ah}a. The
angle consists of two rays and the right one (i.~e., the null geodesic
$x=-t$ from the origin) we shall denote by $\gamma$. The neighbourhood
$U$ of $D^+(\EuScript S)$ is defined to be the union of the left
half-plane and the strip bounded by the graphs of the functions
\[
x(t)= \sqrt{t^2-1}-e^t\quad \text{and}\quad x(t)=e^t-t
\]
The mentioned strip at, say, $t<-2$ is characterized by width $\shir$:
\[
\shir(t)\equiv -\sqrt{t^2-1}-t+2e^t,
\]
which monotonically decreases with $|t|$.

Now pick a sequence of negative numbers such that $t_1<-2$ and
\[
t_{k+1}<t_{k}-1, \quad \shir(t_k)<2^{-k}
\qquad k=1,2,\dots
\]
and define the following portions of the strip:
\[
O_k\equiv\{p\in U\colon\quad x(p)>0,\quad t_{k}-1<t(p)<t_{k}\},\qquad
O\equiv\bigcup_k O_k.
\]

To build the spacetime $M'$, consider a flat strip
\[
\begin{gathered}
R\colon\qquad \rmd s^2= -\rmd \tau^2 +\rmd\chi^2, \\
\tau \in (-1,0),\quad\chi>0
\end{gathered}
\]
and the isometry $\Psi$ which sends, for every $k$, each point $p\in O_k$
to the point $q\in R$ according to the rule\footnote{The recurrent form
of the last formula may be more transparent: $\chi_k=\chi_{k-1}+ (t_{k-1}
- t_k) -\shir_{k-1}$}
\begin{gather*}
\tau(q)= t(p) - t_k,\qquad
\chi(q)= x(p)-\chi_k
\\
\chi_1 \equiv  \sqrt{t_1^2-1}-e^{t_1},\qquad
\chi_k\equiv\chi_1 - t_k +t_1  - \sum_{i=1}^{k-1}\shir_i
\quad\text{at } k>1.
\end{gather*}
 Now $M'$
is defined as the quotient of $U\cup R$ over $\Psi$ [i.~e., as the result
of gluing together $U$ and $R$ by $O$, see Fig.~\ref{fig:Ah}(b)]:
\[
M'\equiv U\cup_\Psi R.
\]
The natural projection $\pi\colon U\to M'$ is an isometric embedding, so
it only remains to prove that   $  \pi(D^+(\EuScript S))\neq
D^+(\pi(\EuScript S))$.

To this end consider  the null  geodesic $\pi(\gamma)$  and denote by
$\EuScript L$ the part of it that lies in $R'\equiv\pi(R)$. $\EuScript L$
is the set of the geodesic segments $\gamma^{k}$ which, in the convenient
coordinates
\[
u(q)\equiv \chi(p) + \tau(p),\quad
 v(q)\equiv \chi(p) - \tau(p),\qquad p\in R,\quad q=R',
\]
are the segments $\gamma^{k}\colon \quad u=u_k$, where
\begin{equation*}
u_k\equiv-t_k - \chi_k=  |t_1|-\chi_1   +
\sum_{i=1}^{k-1}\shir_i,\qquad
k=1,2\dots
\end{equation*}
(at $k=1$ the last term is understood to be zero). Now note that $u_k$
grow with $k$ and converge to some $u_\infty$. Let us check that the
existence of the geodesic $\gamma^{\infty}\colon\quad u=u_\infty$ proves
our claim (in fact, this is obvious from Fig.~\ref{fig:Ah}(b), where
$\gamma^{\infty}$ is shown by the dashed line).

Indeed, any $p\in R'\cap\pi(D^+(\EuScript S))$ belongs to some $O_k$. So,
$u(p)$ must be less than $u_k$ and hence strictly less than $u_\infty$.
Thus $p\notin\gamma^{\infty}$. On the other hand, any past directed
timelike curve $\alpha$ through a point $a\in\gamma^{\infty}$ must have
points with the $u$-coordinate less than $u_\infty$, hence  $\alpha$
meets a $\gamma_k$ in some point $s$. If $s\notin D^+(\pi(\EuScript S))$
(which is, in principle, imaginable, though not, perhaps, in our case),
then the proof is completed, because $s$ is a point of $
\pi(D^+(\EuScript S))$. And if
$s\in D^+(\pi(\EuScript S))$, then $\alpha$ being extended sufficiently
far to the past must meet $\pi( \EuScript S)$, which implies that $a$
(and thus the entire $\gamma^{\infty}$, too) are in $\pi(D^+(\EuScript
S))$.

\end{document}